\documentclass[twocolumn,preprintnumbers,floats,prd,amssymb,floatfix,nofootinbib,balancelastpage,superscriptaddress,amsmath]{revtex4-1}

\pdfoutput=1
\usepackage[utf8]{inputenc}

\usepackage{amssymb}
\usepackage{amsmath}
\usepackage{amsfonts}
\usepackage{graphicx}
\usepackage{color}
\usepackage{xspace}

\usepackage[section]{placeins}
\usepackage{afterpage}

\usepackage{upgreek}

\usepackage{float}
\usepackage{slashed}

\usepackage{multirow,rotating}

\def\gsim{\raise0.3ex\hbox{$\;>$\kern-0.75em\raise-1.1ex\hbox{$\sim\;$}}}
\def\lsim{\raise0.3ex\hbox{$\;<$\kern-0.75em\raise-1.1ex\hbox{$\sim\;$}}}

\newcommand{\ba}[1]{\begin{eqnarray} \label{(#1)}}
\newcommand{\ea}{\end{eqnarray}}

\definecolor{tobycolour}{rgb}{.6,.0,.4}

\definecolor{dcolour}{rgb}{.5, .5, .5}

\newcommand{\AddrAHEP}{
  {\it AHEP} Group, Instituto de F\'{\i}sica Corpuscular --
    CSIC/Universitat de Val{\`e}ncia \\
  Calle Catedr\'{a}tico Jos\'{e} Beltr\'{a}n, 2 E-46980
  Paterna, Spain }

\def\gsim{\raise0.3ex\hbox{$\;>$\kern-0.75em\raise-1.1ex\hbox{$\sim\;$}}}
\def\lsim{\raise0.3ex\hbox{$\;<$\kern-0.75em\raise-1.1ex\hbox{$\sim\;$}}}

\begin{document}

\preprint{\parbox[t]{3.3cm}{APCTP Pre2020-002\\  IFIC/20-01 }}

\title{Heavy neutral leptons at ANUBIS}

\author{Martin Hirsch}
\email{mahirsch@ific.uv.es}
\affiliation{\AddrAHEP}

\author{Zeren Simon Wang}
\email{zerensimon.wang@apctp.org}
\affiliation{Asia Pacific Center for Theoretical Physics (APCTP) 
- Headquarters San 31,\\ Hyoja-dong, Nam-gu, Pohang 790-784, Korea}

\begin{abstract}
Recently Bauer \textit{et al.} \cite{Bauer:2019vqk} proposed ANUBIS, an auxiliary
detector to be installed in one of the shafts above the ATLAS or CMS
interaction point, as a tool to search for long-lived particles. Here,
we study the sensitivity of this proposal for long-lived heavy neutral
leptons (HNLs) in both minimal and extended scenarios. We start with
the minimal HNL model where both production and decay of the HNLs are
mediated by active-sterile neutrino mixing, before studying the case
of right-handed neutrinos in a left-right symmetric model. We then
consider a $U(1)_{B-L}$ extension of the Standard Model (SM). In this model HNLs are
produced from the decays of the mostly SM-like Higgs boson, via mixing
in the scalar sector of the theory. In all cases, we find that ANUBIS
has sensitivity reach comparable to the proposed MATHUSLA detector.
For the minimal HNL scenario, the contributions from $W$'s decaying
to HNLs are more important at ANUBIS than at MATHUSLA, extending the
sensitivity to slightly larger HNL masses at ANUBIS.
\end{abstract}


\vskip10mm

\maketitle
\flushbottom
%
%
\section{Introduction}\label{sec:introduction}

Recent years have seen a surge of interest in long-lived particles
(LLPs), both from experimental and theoretical sides. 
Several new detectors have been proposed to search for LLPs using the
LHC beams: MATHUSLA \cite{Chou:2016lxi,Curtin:2018mvb}, FASER
\cite{Feng:2017uoz,Kling:2018wct,Ariga:2018uku}, CODEX-b
\cite{Gligorov:2017nwh,Aielli:2019ivi} and AL3X
\cite{Gligorov:2018vkc, Dercks:2018wum}.  In addition, there is the
SHiP proposal \cite{Alekhin:2015byh}.  Different from all the other
experiments mentioned above, however, SHiP is a beam-dump experiment.
For a recent review on LLPs at the LHC see Ref.~\cite{Alimena:2019zri}. 
\footnote{A study of potential far detectors at future lepton
	colliders was performed in Ref.~\cite{Wang:2019xvx}.}

Ref.~\cite{Bauer:2019vqk} proposed a new far detector design called
``AN Underground Belayed In-Shaft search experiment'' (ANUBIS) to be
constructed inside one of the shafts above either the ATLAS or CMS
interaction point (IP) at the LHC. In this paper, we estimate
the sensitivity reach of ANUBIS for heavy neutral leptons (HNLs) in
various models and compare it with other proposed LLP far detectors.

From the theory point of view, LLPs appear in a variety of standard
model (SM) extensions. Nowadays, in the literature very often dark
matter motivated ``Higgs portal'' models are discussed as a motivation
for LLPs. However, LLPs have a long history. For example several
supersymmetric models, such as gauge mediated Supersymmetry (SUSY)
breaking models and split SUSY, have been known for quite some time to
predict LLPs; for a recent review see Ref.~\cite{Lee:2018pag}. We
would also like to mention explicitly bilinear R-parity breaking SUSY,
where it has been shown that the lifetime of the LLP is correlated
with the small neutrino masses \cite{Porod:2000hv}.

The observation of neutrino oscillations has firmly established that
neutrinos have non-zero mass. (For the experimental status of neutrino
data, see for example Refs.~\cite{deSalas:2017kay,deSalas:2018bym}.)  Many
different models have been proposed to understand the small neutrino
masses. The simplest is the well-known type-I seesaw
\cite{Minkowski:1977sc,Yanagida:1979as,Mohapatra:1979ia,GellMann:1980vs,Schechter:1980gr}.
Here, neutral leptons are added to the SM particle
content. These HNLs mix with the ordinary neutrinos after electroweak
symmetry breaking, typically $V_{\alpha N} \propto (Y^{\nu}_{\alpha
	N}v)/m_N$, with the light neutrino mass given by the famous seesaw
relation: $m_{\nu}\propto (Y^{\nu}v)^2/m_N$. The decay width of these
HNLs are suppressed by $|V_{\alpha N}|^2$, leading automatically to
very long-lived HNLs, if their mass is below the electroweak scale.
\footnote{The mixing between sterile and active neutrinos can be
	larger in non-minimal seesaw models such as the inverse seesaw
	\cite{Mohapatra:1986bd}.}

In the simplest HNL model both production cross section and decay
width are proportional to $|V_{\alpha N}|^2$. However, there are a
number of SM extensions, which give much larger HNL production cross
sections, while still maintaining very long decay lengths.  We will
estimate ANUBIS's sensitivity for two of these: (i) the minimal
left-right symmetric model (LRSM)
\cite{Mohapatra:1974gc,Pati:1974yy,Mohapatra:1980yp} and (ii) the SM
extended by an additional $U(1)_{B-L}$ group. For the latter we use
the model variant discussed in some other recent papers studying LLPs
\cite{Deppisch:2018eth,Deppisch:2019kvs,Amrith:2018yfb}. In all three
model variants we find that ANUBIS has sensitivity reach comparable to
the proposed MATHUSLA detector.

In the following section, we will introduce the different models with
HNLs. Sec.~\ref{sec:simulation} is devoted to describing the detector
setup and the details of the simulation. In Sec.~\ref{sec:results} we
present the numerical results. We conclude and summarize our findings
in Sec.~\ref{sec:conclusions}.

\section{Model basics}\label{sec:models}

\subsection{The minimal HNL scenario}

The minimal HNL model adds $n$ species of HNLs to the particle content
of the SM. They enter both charged-current (CC) and neutral-current
(NC) interactions, suppressed compared to the SM electroweak coupling
strength by small mixing parameters:
\begin{eqnarray}\label{CC-NC}
	{\cal L} &=& \frac{g}{\sqrt{2}}\, 
	V_{\alpha N_j}\ \bar \ell_\alpha \gamma^{\mu} P_L N_{j} W^-_{L \mu} 
	+\nonumber \\
	&&\frac{g}{2 \cos\theta_W}\ \sum_{\alpha, i, j}V^{L}_{\alpha i} V_{\alpha N_j}^*  
	\overline{N_{j}} \gamma^{\mu} P_L \nu_{i} Z_{\mu},
\end{eqnarray}
where $i=1,2,3$, $j=1, .., n$, and $\ell_\alpha$
($\alpha=e,\,\mu,\,\uptau$) are the charged leptons of the SM.
$V_{\alpha N_j}$ labels the mixing between SM light neutrinos and the
HNLs of mass $m_{N_j}$. By $V^{L}_{\alpha i}$ we denote the mixing
within the active neutrino sector, in the basis where the charged
lepton mass matrix is diagonal. With this choice of basis, the
elements of $V^{L}_{\alpha i}$ corresponds to the mixing angles
measured in oscillation experiments.

Neutrino data requires that $n \ge 2$. However, for simplicity we
assume that there is only one species of the HNL, $N$, light enough to
be produced at the LHC. We treat $V_{\alpha N}$ as a free parameter,
to be determined experimentally, and calculate the total decay width
of the HNL using the analytical formulas given in
Ref.~\cite{Atre:2009rg}. Such HNLs are most dominantly produced from
on-shell decays of various kinds of SM particles at the LHC. For this
minimal HNL scenario, we take into account different production
channels: $D-$mesons, $B-$mesons, $W-$bosons, $Z-$bosons, SM Higgs
bosons $h$, and the top quark $t$.  For the mesons channels, we
include into the calculation both three-body and two-body decays. Note that there is a recent study on the sensitivity of future experiments for three-body decays of mesons \cite{Chun:2019nwi}.

\subsection{The minimal left-right symmetric model}

We consider the minimal left-right symmetric model which has gauge
group $SU(3)_C\times SU(2)_L \times SU(2)_R \times U(1)_{B-L}$
\cite{Mohapatra:1974gc,Pati:1974yy}. Right-handed neutrinos are
necessarily present in this model in the right-handed lepton
doublet. Breaking the left-right symmetry group to the SM group via a right
triplet scalar generates Majorana masses for the right-handed
neutrinos and thus automatically a seesaw mechanism
\cite{Mohapatra:1980yp}.

The CC and NC interactions relevant for our analysis are
\begin{eqnarray}\label{CC-NC-LR}
	{\cal L} &=& \frac{g_{R}}{\sqrt{2}}  
	\left(\bar{d}  \gamma^{\mu} P_R u 
	+  V^{R}_{\alpha N}\cdot \bar l_{\alpha} \gamma^{\mu} P_R N\  \right) W^-_{R \mu} 
	+  \\
	\nonumber
	&+&\frac{g_{R}}{\sqrt{1-\tan^{2}\theta_{W} (g_{L}/g_{R})^{2}}}\times\\
	\nonumber
	&& Z_{LR}^{\mu}\bar{f} \gamma_{\mu}\left[ T_{3R} 
	+  \tan^{2}\theta_{W} (g_{L}/g_{R})^{2}\left(T_{3L} - Q\right)\right] f  
\end{eqnarray}
where $V^{R}_{\alpha N}$ is the right-handed sector neutrino mixing matrix.  
Different from $V_{\alpha N}$ describing left-right mixing, one 
expects that the entries of  $V^{R}_{\alpha N}$ are ${\cal O}(1)$. 
The charged $W_{L,R}$-boson states can be expressed in terms of the mass
eigenstates, by a mixing angle $\zeta$. Since $\zeta \ll 1$, 
for simplicity we will call the $W$-bosons $W_{L,R}$, instead of 
the more correct (but somewhat awkward) $W_{1,2}$. 

For the production of the right-handed neutrinos at the LHC in this
model, there are different contributions. First, there is the on-shell
production of $W_R$, with the $W_R$ decaying to $N + l_{\alpha}$.
Second, $N$ can appear in the decays of mesons, through the decays of
off-shell $W_L$, suppressed by the small active-sterile neutrino
mixing $V_{\alpha N}$. And, finally, mesons can decay via off-shell
$W_R$ to $N$ plus charged leptons\footnote{See Ref.~\cite{Mandal:2017tab} for an earlier study on the sensitivity of meson decays in the LRSM at a number of future searches.}. Relative to the off-shell $W_L$
contribution, the off-shell $W_R$ contribution then can be estimated
via $|V_{\alpha N}|^2\rightarrow |V^{R}_{\alpha N}|^2\Big( g_{R}/g_{L}
\Big)^4 \Big( m_{W_L}/m_{W_R} \Big)^4$.  For our sensitivity
estimates, we assume that $|V_{\alpha N}|^2$ is given by the naive
seesaw estimate and, thus, small enough that the $N$ production from
meson decays and the decay width of $N$ are dominated by the off-shell
$W_R$ contribution. Note, that contributions to $N$ decay from
off-shell $Z_R$ are completely negligible in this approximation.

We will again focus on the case with only one right-handed neutrino
$N$ within the kinematic region of interest ($m_N \lsim m_B$).
We will use $g_L/g_R=1$ in our numerical study, the results can 
be easily scaled to other values of $g_R$.

\subsection{HNLs in $U(1)_{B-L}$}

One of the simplest SM extensions automatically predicting HNLs is
adding a gauged $U(1)_{B-L}$ to the SM gauge group
\cite{Davidson:1978pm,Mohapatra:1980qe}. The model variant we will be using in our
numerical estimate and its implications for accelerator searches has
been discussed recently in Refs.~
\cite{Deppisch:2019kvs,Accomando:2016rpc,Deppisch:2018eth,Das:2018tbd,Das:2019fee}.  Apart
from the HNLs, the model predicts a $Z'$ and (at least) a second
Higgs. This new scalar, required to break $U(1)_{B-L}$, will mix with
the (mostly) SM Higgs, observed at the LHC, via a small mixing angle
$\beta$.

Two new channels for producing HNLs, besides the standard
active-sterile neutrino mixing, then exist in this model. First, HNLs
can be pair produced from a light $Z'$ decay \cite{Deppisch:2019kvs},
or, second, the SM-like Higgs will decay through mixing to a pair of HNLs
\cite{Accomando:2016rpc,Deppisch:2018eth}. In our numerical
calculation, we choose to concentrate on the SM-like Higgs decay
channel. The decay width of the HNLs to SM particles
in this model again is proportional to the mixing $|V_{\alpha N}|^2$.

We choose the benchmark values for the model parameters, following
the discussion and choices in
Refs.~\cite{Deppisch:2018eth,Chiang:2019ajm}:
\begin{eqnarray}
	m_N &=& 1-60 \text{ GeV}, V_{\alpha N}=10^{-9}-10^{-2},\nonumber\\
	m_{Z'}&=&6 \text{ TeV}, g'_1=0.8, \tilde{x}=3.75 \text{ TeV},
	\label{eqn:U1BmLHiggsParameters}\\
	&&m_{H}=450 \text{ GeV}, \sin{\beta}=0.3, \nonumber
\end{eqnarray}
where $\tilde{x}=m_{Z'}/2g'_1$ is the VEV of the extra scalar, $\beta$
is the scalar mixing angle, and $m_H$ is the mass of the heavier
scalar. (The latter does not affect our calculation
directly.) Ref.~\cite{Chiang:2019ajm} derived current lower (upper)
limits on $m_{Z'}$ ($g'_1$) by recasting the ATLAS and CMS searches
\cite{Sirunyan:2018exx,Aad:2019fac}. As for the latest bounds on $m_H$
and $\sin{\beta}$, one may refer to Ref.~\cite{Amrith:2018yfb}.

The decay width of the SM-like Higgs into a pair of HNLs for one single
generation is \cite{Accomando:2016rpc}:
\begin{eqnarray}
	\Gamma(h\rightarrow NN) = \frac{1}{2}\frac{m_N^2}{\tilde{x}^2}\sin^2{\beta}\,\frac{m_h}{8\pi}\,\Big( 1- \frac{4 m_N^2}{m_h^2} \Big)^{3/2},
\end{eqnarray}
where we assume that among the three generations of HNLs predicted,
only one generation labelled with $N$ is below the Higgs decay
threshold: $m_{N}<m_h/2$.

The decay branching ratio of the SM-like Higgs bosons into a pair of HNLs 
can then be written as:
\begin{eqnarray}\label{eqn:Brh2NN}
	\text{Br}(h\rightarrow NN) = \frac{\Gamma(h\rightarrow NN)}{\Gamma(h\rightarrow NN) + \cos^2{\beta}\,\Gamma^h_{\text{SM}}},
\end{eqnarray}
where $\Gamma^h_{\text{SM}}=4.1$ MeV \cite{CERNHiggsWidth} is the SM
Higgs total decay width for $m_h=125.10$ GeV.

Production cross section for a pair of $N$'s is expressed with the
following formula:
\begin{eqnarray}
	\sigma(pp\rightarrow h \rightarrow N N) = \cos^2{\beta}\cdot  \sigma(pp\rightarrow h)_{\text{SM}}\cdot \text{Br}(h\rightarrow NN),\nonumber\\
\end{eqnarray}
where the SM production cross section of the Higgs boson is modified
by a factor $\cos^2{\beta}$ and $\text{Br}(h\rightarrow NN)$ is
calculated with Eq.~\eqref{eqn:Brh2NN}.

\section{Detector setup \& Simulation}\label{sec:simulation}

\begin{figure}
	\includegraphics[scale=0.5]{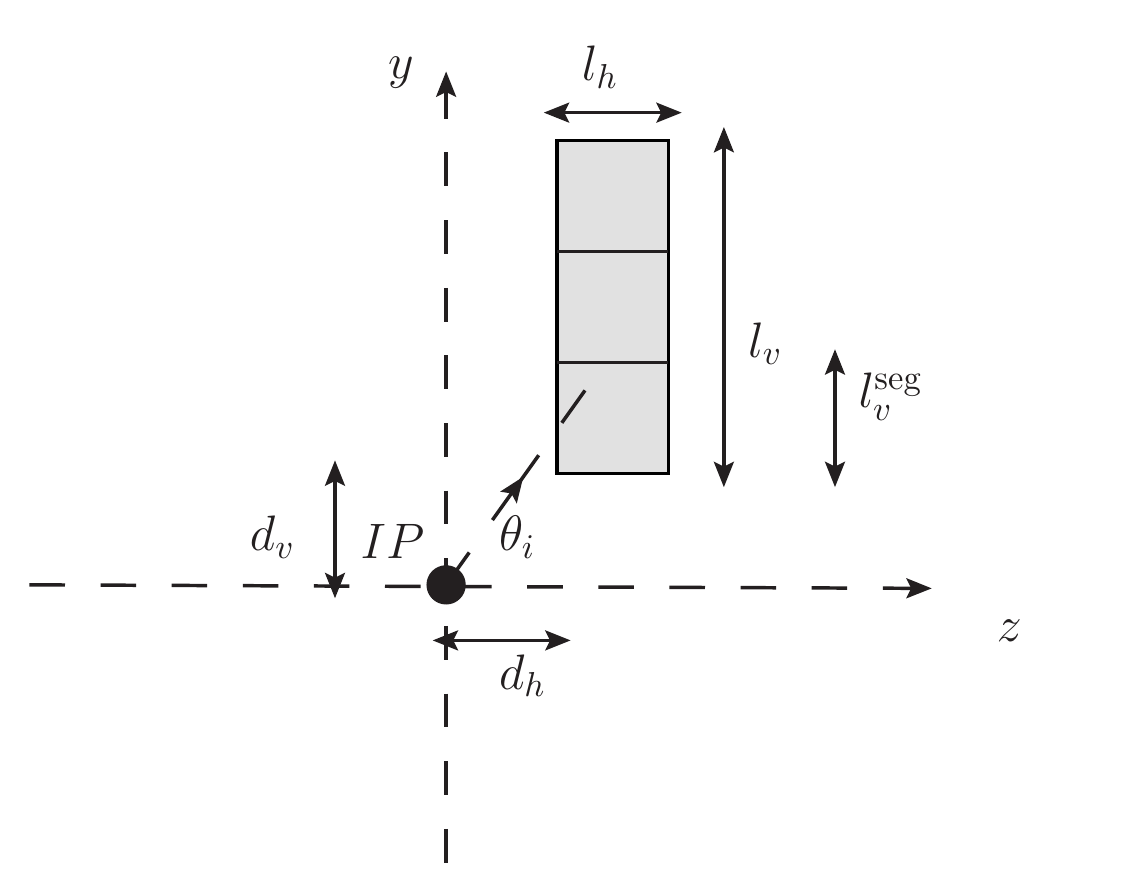}
	\includegraphics[scale=0.5]{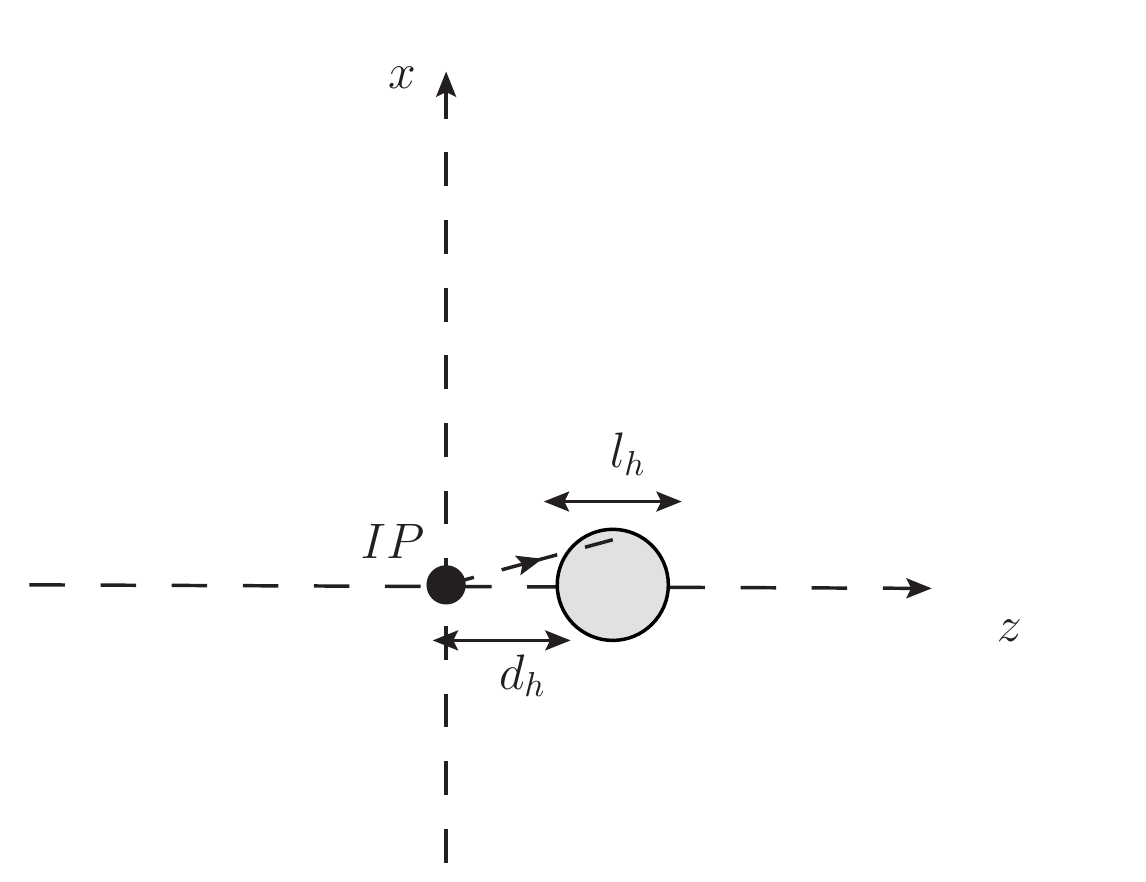}
	\caption{Profile sketches of the ANUBIS detector in the $y-z$
		and $x-z$ planes, respectively. In the upper plot we depict
		a sample long-lived HNL by the arrowed dashed line with
		$\theta_i$ labeling its polar angle.}
	\label{fig:detectorsketch}
\end{figure}

Fig.~\ref{fig:detectorsketch} shows two profile sketches of the ANUBIS
detector setup in the $y-z$ and $x-z$ planes, respectively, where the
collider beams are oriented along the $z-$direction. The cylindrical
detector is horizontally (vertically) displaced from the IP with a
distance $d_h=5$ ($d_v=$24) m. It has a length of $l_v=56$ m and a
diameter $l_h=18$ m. With four tracking stations, the detector may be
divided into three segments with the length of each segment
$l_v^{\text{seg}}$ roughly equal to 18.7 m. We label the polar angle
of a sample long-lived $N_i$ with $\theta_i$.

\begin{table}[]
	\begin{center}
		\begin{tabular}{|c|c|c|c|c|c|c|}
			\hline
			$M$ & $b\bar{b}$ & $c\bar{c}$ & $h$ & $t\bar{t}$ & $W$ & $Z$\\
			\hline
			$\sigma_M$ [pb] &$6 \times 10^8$ & $2 \times 10^{10}$& 
			$6\times 10^1$ & $1\times 10^3$ & $2.1\times 10^5$ &  $6.4\times 10^4$\\		
			\hline
		\end{tabular}
		\caption{(Rough) production cross section of each type of mother particles
			of the HNLs at LHC with $\sqrt{s}=14$ TeV. For $c\bar{c}$ we take
			into account all of $D^0$, $D^+$, $D_s$, and $D^{0*}$ mesons. For
			references see text.}
		\label{tab:diffchannels}
	\end{center}
\end{table}

The total number of HNLs produced is calculated with
\begin{eqnarray}
	n_N = n_M \cdot \text{Br}(M\rightarrow n\, N)\cdot n,
\end{eqnarray}
where $n_M$ stands for the total number of a mother particle $M$
produced.  For high-luminosity LHC's integrated luminosity
$\mathcal{L}_{\text{int}}=3$ ab$^{-1}$ and the center-of-mass energy
$\sqrt{s}=14$ TeV, $n_M=\mathcal{L}_{\text{int}} \cdot \sigma_M$ for
$M=h$, $W,$ and $Z$, and twice that for $M$ being $D-$mesons,
$B-$mesons, and top quarks, since the latter are mainly pair-produced
at the LHC.  $\sigma_M$ denotes the production cross section of each
channel, and $\text{Br}(M\rightarrow n\, N)$ labels the decay
branching ratio of $M$ into $n$ $N$'s ($n=1$, 2, $\ldots$).  In
Table~\ref{tab:diffchannels}, we list the production cross section
over a whole sphere for each channel at the LHC for $\sqrt{s}=14$ TeV.
We follow the procedure explained in Ref.~\cite{Helo:2018qej} to
obtain $\sigma_M$ for $D-$, $B-$mesons, $W-$bosons, and $Z-$bosons at
$\sqrt{s}=13$ TeV.  We then rely on the Monte-Carlo (MC) simulation
tool \texttt{Pythia 8.243} \cite{Sjostrand:2006za,Sjostrand:2007gs} to
extrapolate the cross section to $\sqrt{s}=14$ TeV. Slightly depending
on the production channel, cross sections are changed by roughly a
factor $\sim 1.07$ by this change of $\sqrt{s}$.  As for the
production cross section of $h$ and $t\bar{t}$, we extract the numbers
from Refs.~\cite{HiggsXSec14, ttbarXSec14}.

We express the total number of visible displaced vertices inside the
ANUBIS as a function of $n_N$, average decay probability inside the
fiducial volume $\langle P[N \text{ in f.v.}] \rangle$, and the decay
branching ratio of the HNLs into visible states (we include all the
decay channels except the completely invisible tri-neutrino one):
\begin{eqnarray}
	n^{\text{vis}}_N = n_N \cdot \langle P[N \text{ in f.v.}] \rangle \cdot \text{Br}(N\rightarrow\text{visible}),
\end{eqnarray}
where $\langle P[N \text{ in f.v.}] \rangle$ is estimated by performing MC simulation with \texttt{Pythia 8}:
\begin{eqnarray}
	\langle P[N \text{ in f.v.}] \rangle = \frac{1}{n_{N}^\text{MC}}\sum_{i=1}^{n_N^\text{MC}}P[N_i\text{ in f.v.}].
\end{eqnarray}
$n_{N}^{\text{MC}}$ labels the total number of MC-simulated $N$'s and
$P[N_i\text{ in f.v.}]$ denotes the individual decay probability of
each simulated $N$ taking into account the detector geometry of
ANUBIS and the kinematics of the HNLs.  To evaluate $P[N_i\text{ in
	f.v.}]$, we divide the ANUBIS decay volume into three segments
corresponding to its 4 equally spaced tracking stations and add up the
decay probabilities inside each segment:
\begin{align}
	&P[N_i\text{ in f.v.}]=\sum_{j=1}^{3}\,\frac{\delta\phi^j}{2\pi}\,\cdot  e^{\frac{-L_i^j}{\lambda_i^z}} \cdot (1-e^{-\frac{L^{j'}_i}{\lambda_i^z}})\,,
	\label{eqn:decayprobability}\\
	\delta\phi^j&=2\arctan{\frac{l_h/2}{d_v+(2j-1)/2\cdot l_v^{\text{seg}}}},\\
	L_i^j&=\text{min}\bigg(\text{max}\bigg(d_h,\frac{d_v+(j-1)\cdot l_v^\text{seg}}{\tan{\theta_i}}\bigg),d_h+l_h\bigg)\,,\\
	L_i^{j'}&=\text{min}\bigg(\text{max}\bigg(d_h,\frac{d_v+j\cdot l_v^{\text{seg}}}{\tan{\theta_i}}\bigg),d_h+l_h\bigg)-L_i^j\,,\\
	\lambda_i^z&=\beta^z_i \, \gamma_i\,  c \, \tau\,.\label{eqn:boosteddecaylength}
\end{align}
Eq.~\eqref{eqn:boosteddecaylength} calculates the boosted decay length
of the $i-$th simulated $N$ along the beam direction, making use of
its speed in the longitudinal direction $\beta^z_i$, boost factor
$\gamma$, and proper lifetime $\tau$ with $c$ denoting the speed of
light.  The speed and boost factor can be easily obtained from
kinematic information of the HNLs provided in \texttt{Pythia 8}:
\begin{eqnarray}
	\beta^z_i = |p^z_i/E|,\, \gamma_i = E_i/m_N,
\end{eqnarray}
where $p_i^z$ is the momentum along the $z-$direction, and $E_i$ is the
energy.  For the azimuthal angle coverage we choose the middle
position of each segment in order to fix the reference height.

We simulate 100k events for each parameter point using \texttt{Pythia
	8}. In order to obtain the maximal number of statistics, we require
the mother particles to exclusively decay in the channels leading to
HNL production. We then calculate the decay branching ratios of the
mother particles into HNLs analytically.

We cross-checked the usage of Eq.~\eqref{eqn:decayprobability} by
including $N$'s decay in \texttt{Pythia 8} and simulating 10 million
events and compared the sensitivity plots of both approaches.  This
method is very time-consuming and does not lead to smooth sensitivity
curves even with 10 million MC events. However, overall both 
calculations lead to similar final results.

\section{Results}\label{sec:results}

We present numerical results in this section.
As discussed in Ref.~\cite{Bauer:2019vqk}, ANUBIS should be a nearly background-free experiment.
Backgrounds from cosmic rays should be negligible as a result of directional cuts.
A more important background source could be long-lived neutral SM particles, such as $K^0_L$, which might reach ANUBIS and decay within the detector volume.
Ref.~\cite{Bauer:2019vqk} proposes to use the ATLAS calorimeter as an active veto to eliminate this background.
In our figures below we always show the 3-events sensitivity lines for ANUBIS. In case zero background can be reached, these lines correspond to 95\% C.L. (confidence level) limits.

\subsection{Sensitivities for different production modes}

\begin{figure}
	\includegraphics[scale=0.65]{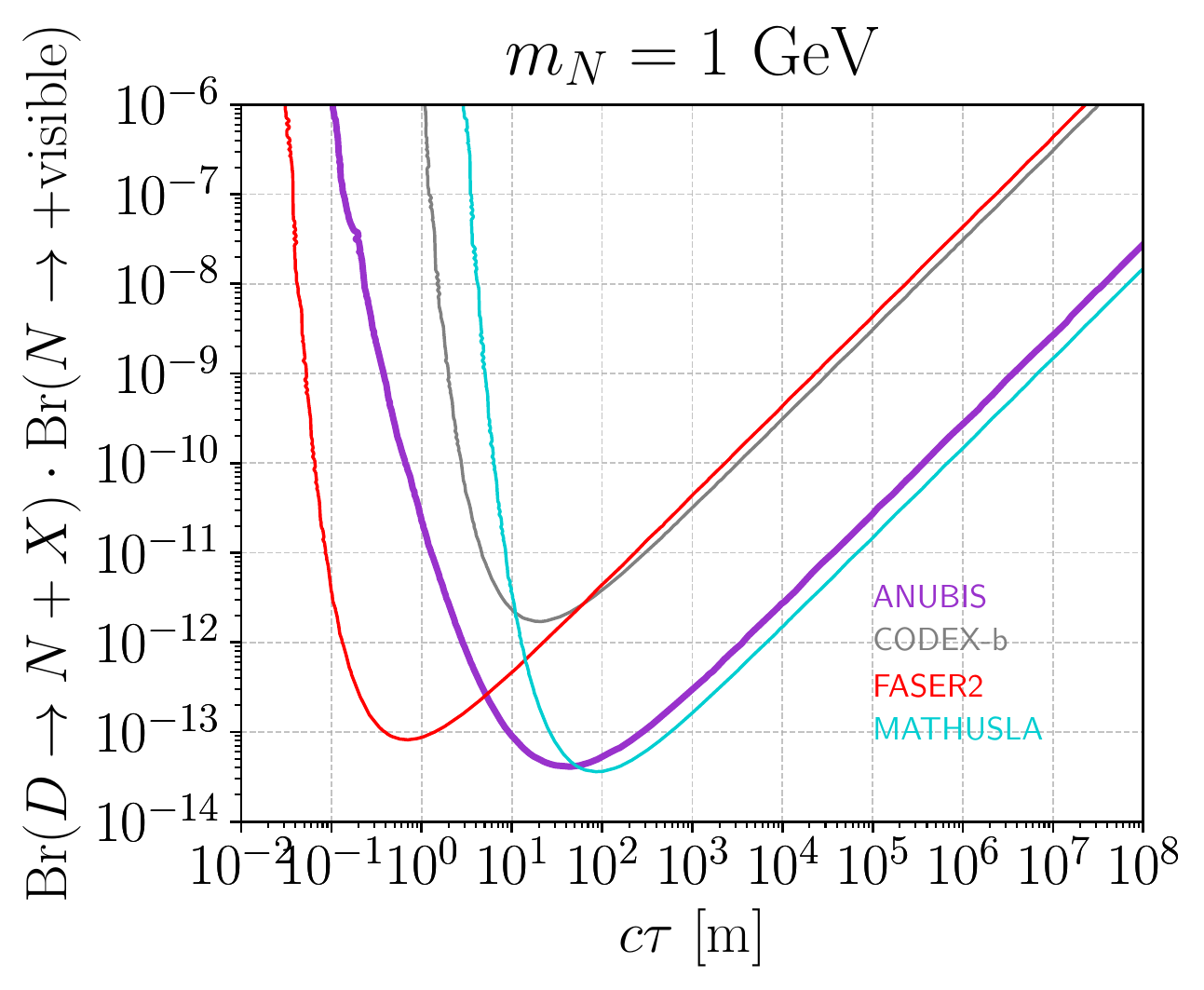}
	\includegraphics[scale=0.65]{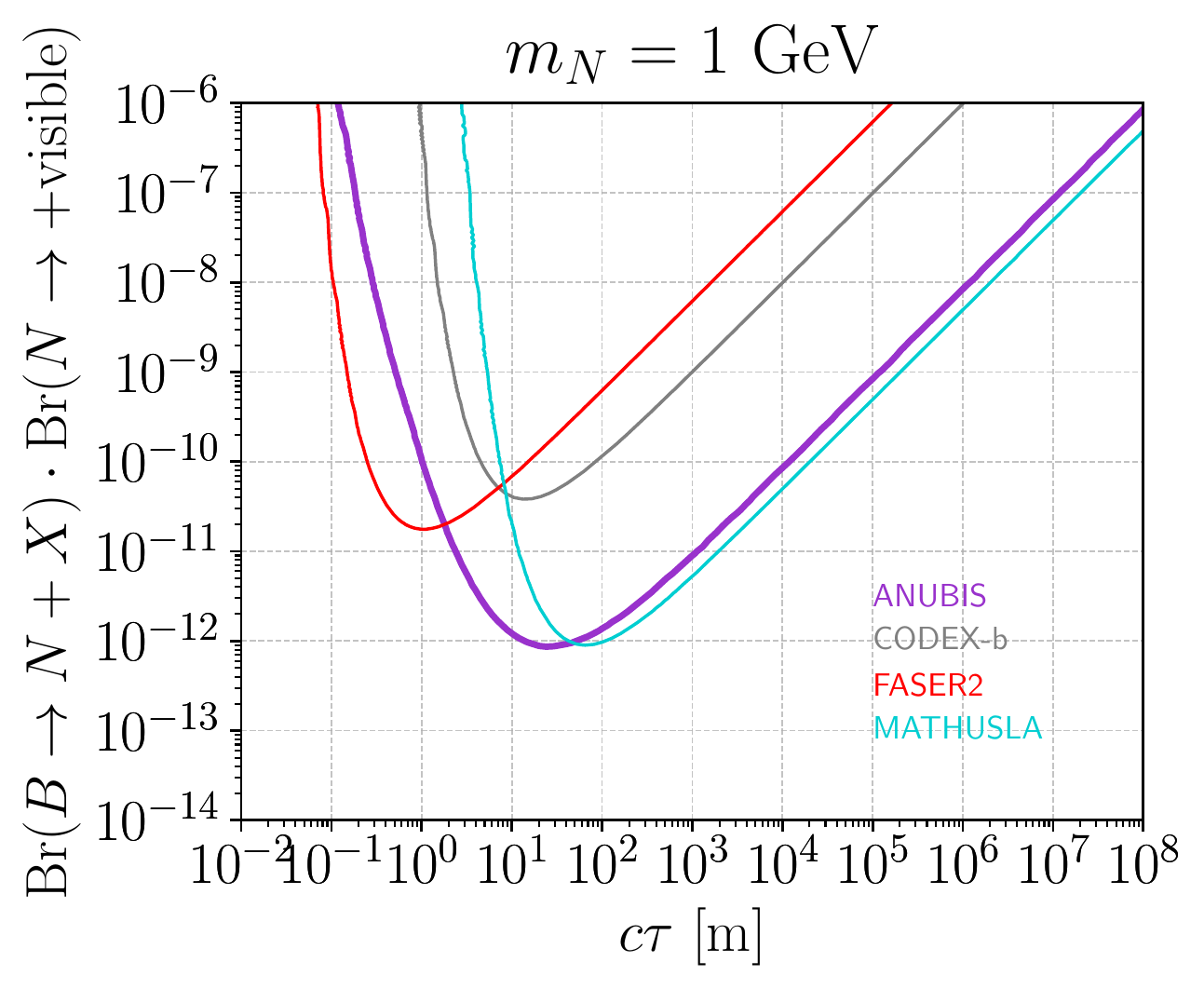}
	\caption{The sensitivity estimates for ANUBIS, compared with
		MATHUSLA, CODEX-b and FASER2, in the plane branching ratio
		versus $c\tau$, for HNLs from $D$-decays (upper plot) and
		$B$-decays (lower plot), in the context of the minimal HNL
		scenario.}
	\label{fig:HNL_Br_vs_ctau}
\end{figure}

In Fig.~\ref{fig:HNL_Br_vs_ctau} we present sensitivity estimates for
ANUBIS for a long-lived neutral fermion of mass 1 GeV, produced from
$D-$ and $B-$mesons decays, using the minimal HNLs as the benchmark
model. The two plots in the figure show the plane Br$(D/B\rightarrow N
+ X)\cdot$Br$(N\rightarrow \text{visible})$ vs. $c\tau$, where $c\tau$
is the proper decay length of $N$, and Br$(N\rightarrow
\text{visible})\approx 91.2\%$ for $m_N=1$ GeV.  

The exclusion-limit isocurves for CODEX-b (300 fb$^{-1}$), FASER2 (3 ab$^{-1}$), and MATHUSLA (3 ab$^{-1}$) are
reproduced from Ref.~\cite{Helo:2018qej}\footnote{The setup of
	``FASER$^{R}$'' considered in Ref.~\cite{Helo:2018qej} is only
	slightly different from the approved configuration of FASER2
	described in Ref.~\cite{Ariga:2019ufm}, and the final results should
	be similar. During the completion of this work, we noticed that when
	we produced the plots in Fig.~1 of Ref.~\cite{Helo:2018qej}, we
	mistakenly failed to absorb the phase space suppression factor for a
	massive HNL into Br($B\,(D)\rightarrow N +X$), leading to a
	reduction of a factor $\sim$2 (10) in Br($B\,(D)\rightarrow N +X$)
	reach for the left (right) plot of Fig.~1 in
	Ref.~\cite{Helo:2018qej}.  This computational mistake has been
	corrected here and the curves are updated in
	Fig.~\ref{fig:HNL_Br_vs_ctau}. The final results of Ref.~
	\cite{Helo:2018qej} are unaffected by this error.}.  As the figure
shows, for both $B$- and $D$-decays the minimum branching ratios that
can be explored are very similar for ANUBIS and MATHUSLA, despite
ANUBIS having a much smaller instrumented volume.  This can be simply
traced to the much smaller distance of ANUBIS from the IP.  As a
result of this and of the fact that the HNLs of mass 1 GeV traveling
inside the window of MATHUSLA typically have boost factors larger than
those of the HNLs traveling towards ANUBIS by less than a factor 2,
relative to ANUBIS, MATHUSLA's maximal sensitivity occurs at larger
values of $c\tau$.  Compared to FASER2 and CODEX-b, ANUBIS shows
clearly better sensitivity.  Note that the huge change in FASER2's
sensitivity relative to the other experiments, when going from $B$- to
$D$- meson decays. This is caused by the very forward position of
FASER2, with lighter particles produced more forward at the
LHC. Recall, that CODEX-b is supposed to have only $10^3$ m$^3$ of
decay volume, with FASER being even smaller, while ANUBIS will have
roughly $1.5 \times 10^4$ m$^3$ of volume.

We also comment on the usage of the mean value of $\beta\cdot\gamma$
for calculating the decay probability in Ref.~\cite{Helo:2018qej}.
For large $c\tau$ in the linear regime, this choice is a good
approximation.  However, for small values of $c\tau$, the tail of the
$\beta\cdot\gamma$ distribution has a dominant contribution to the
decay probability, and thus using the average $\beta\cdot\gamma$
underestimates the reach in the decay branching ratios of the mesons
into HNLs.

\subsection{Results- the minimal scenario and the LRSM}

\begin{figure}
	\includegraphics[scale=0.65]{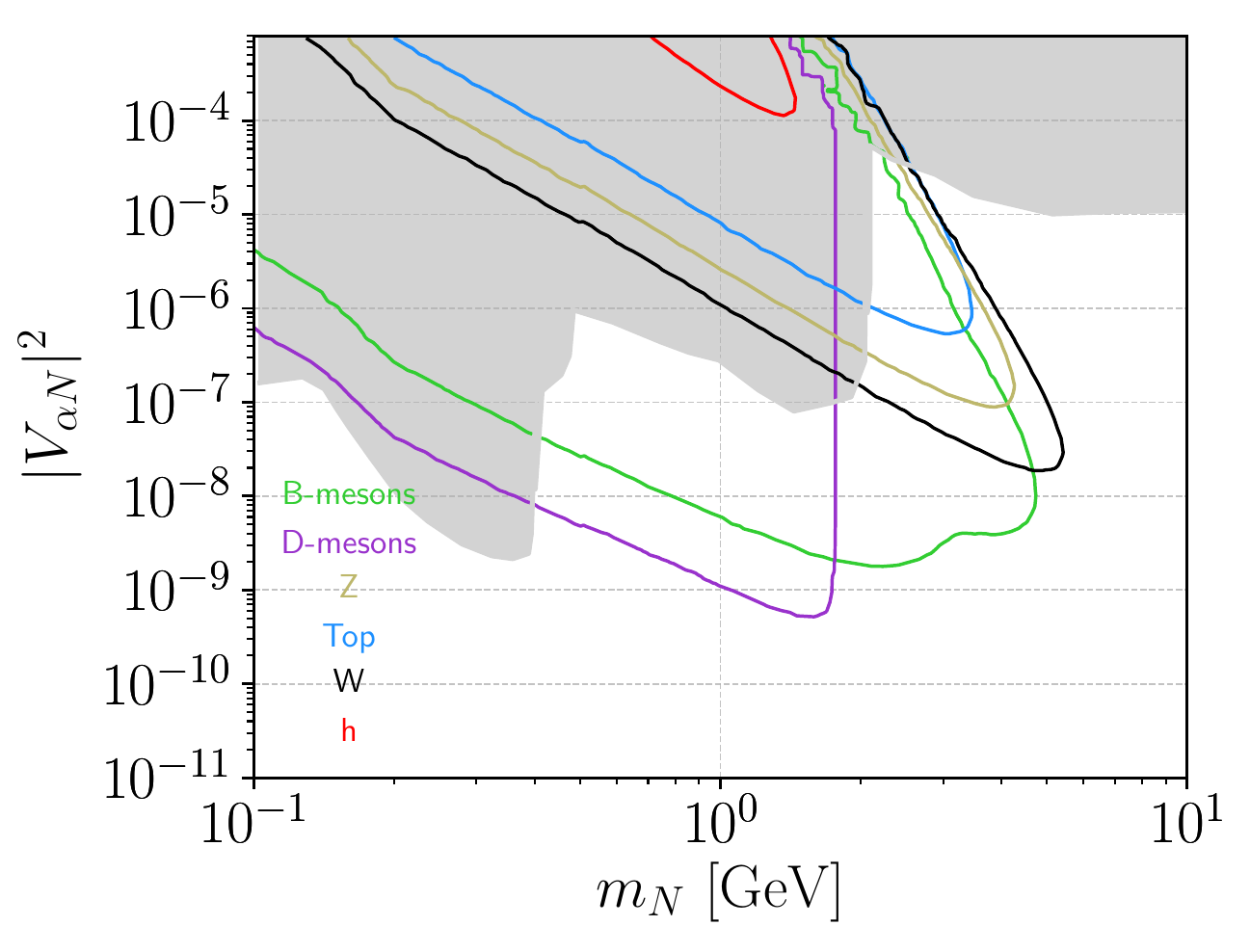}
	\includegraphics[scale=0.65]{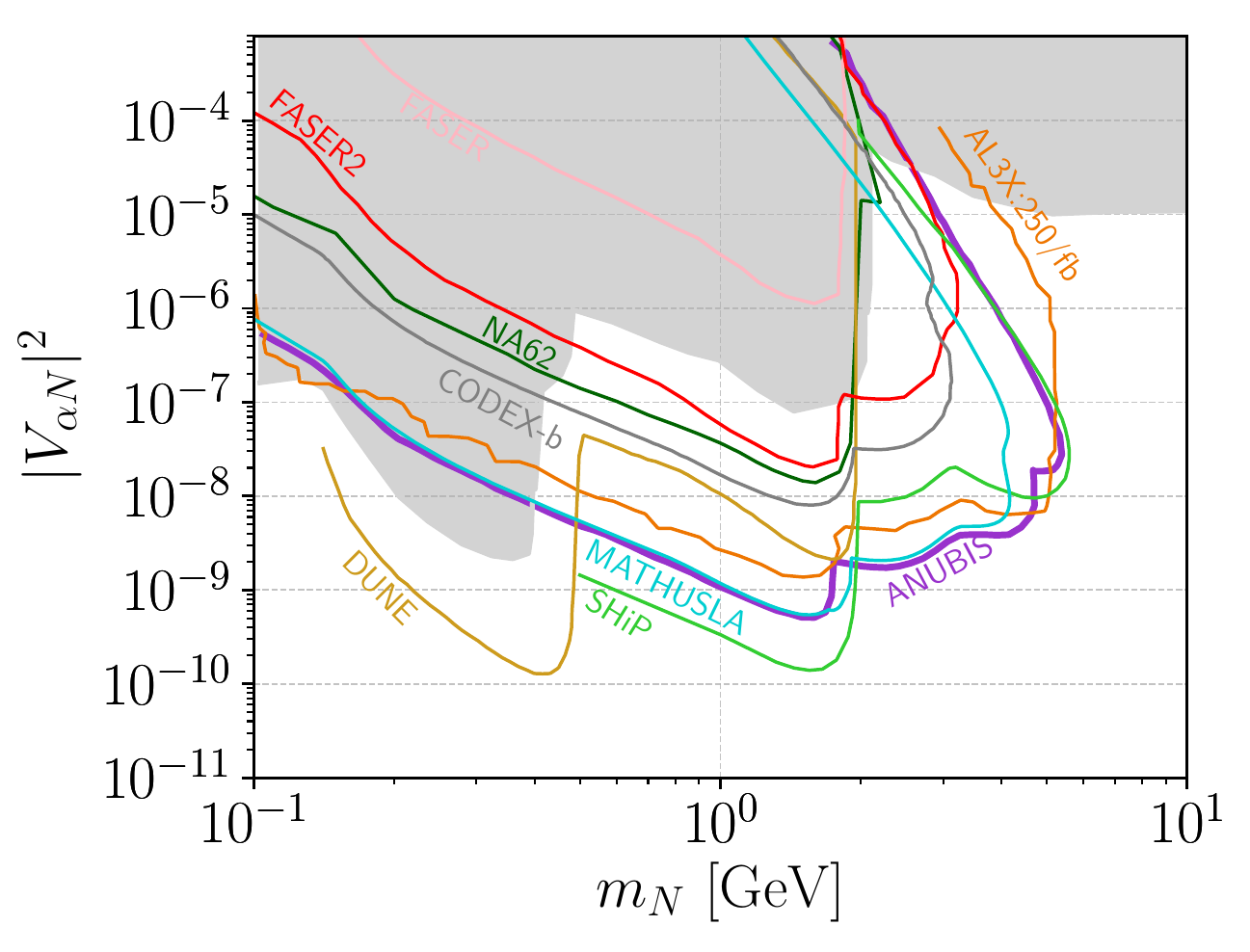}
	\caption{The sensitivity reach of ANUBIS for HNLs produced
		from different channels (upper figure) and reach compared to
		other future experiments (lower figure), in the context of the
		minimal HNL scenario, with one generation of $N$ mixing with
		$\nu_\alpha$, $\alpha=e/\mu$.}
	\label{fig:HNL_sensitivities_minimal}
\end{figure}

We show the results for the minimal model of HNLs in
Fig.~\ref{fig:HNL_sensitivities_minimal} in the plane $|V_{\alpha
  N}|^2$ vs. $m_N$, where we consider one generation of $N$ mixing
with either $\nu_e$ or $\nu_\mu$ but not both ($\alpha=e/\mu$).  The
upper plot compares the exclusion limits among different production
channels at ANUBIS, while the lower compares combined sensitivities of
the dominant production modes: $B-$, $D-$mesons, and $W-$bosons at
ANUBIS with the limits at other future experiments.  We show in the
gray area in the background the experimentally excluded region of the
parameter space based on the summary given in
Ref.~\cite{Deppisch:2015qwa}, including the searches from CHARM
\cite{Bergsma:1985is}, PS191 \cite{Bernardi:1987ek}, JINR
\cite{Baranov:1992vq}, and DELPHI \cite{Abreu:1996pa}.  For the other
future experiments, we extract the sensitivity projections from a
series of past studies
\cite{Krasnov:2019kdc,Drewes:2018gkc,Bondarenko:2018ptm,Dercks:2018wum,Ariga:2018uku,Helo:2018qej}.
Note that for the sensitivity prediction of DUNE extracted from Ref.~\cite{Krasnov:2019kdc}, we only show the results for the HNLs mixing with the electron neutrinos. For the muon mixing
  case, the sensitivity limits only get reduced at roughly two mass
  thresholds $m_K-m_\mu\sim 400$ MeV and $m_D - m_\mu \sim 1800$ MeV,
  compared to the case where the electron mixing is
  dominant. Therefore, in order to keep the plot clean, we refrain
  from showing the muon mixing sensitivity curve for DUNE.

The comparison among different production modes shows that the heavy
meson decays into HNLs have the strongest reach in $|V_{\alpha N}|^2$
at $\mathcal{O}(10^{-9})$ ($5\times 10^{-10}$) for $m_N \lsim 4$
($\sim 1.7$) GeV, and that the $W-$channel ($W^+\rightarrow
e^+/\mu^+\, N$) extends the mass reach to almost $m_N=$6 GeV for
mixing squared of $\sim 2\times 10^{-8}$.  Compared to these channels,
the Higgs bosons ($h\rightarrow N \nu_\alpha$), top quarks
($t\rightarrow W^+ b,\, W^+\rightarrow e^+/\mu^+\, N$), and $Z-$bosons
($Z\rightarrow N \nu_\alpha$) would have more limited contributions.

The lower plot of Fig.~\ref{fig:HNL_sensitivities_minimal} compares
the different future experiments' sensitivities on minimal HNLs.  The
exclusion limits of ANUBIS are comparable to that of MATHUSLA for $m_N
\lsim 4$ GeV, and show the advantage for $m_N$ slightly larger than 4
GeV by virtue of its better acceptance for HNLs produced from
$W-$bosons decays, where MATHUSLA loses sensitivity.

The plots shown in Fig.~\ref{fig:HNL_sensitivities_minimal} are valid for $\alpha=e,\mu$, assuming that the efficiencies for electrons and muons in ANUBIS are at least approximately equal.
We also want to comment briefly on possible constraints on $V_{\tau N}$. Ref.~\cite{Bauer:2019vqk} does not contain any information on detection efficiencies for $\tau$'s.
Thus, we can not give definite predictions for $V_{\tau N}$.
However, a few qualitative comments might be in order.
Both of ANUBIS and MATHUSLA are sparsely instrumented, large-volume tracking detectors.
In both experiments, muons will show up as single tracks, while taus will give either one or three collimated tracks.
A discussion of tau detection in MATHUSLA can be found in Ref.~\cite{Curtin:2017izq}.
The HNLs sensitivity curves for $V_{\mu N}$ and $V_{\tau N}$ shown in Ref.~\cite{Curtin:2018mvb} allow us to estimate the sensitivity loss for MATHUSLA for the case of taus, to be in the order of $(30-50)$ for mixing squared, depending on HNL mass.
Given the similarities in the detector principles, it may not be unreasonable to suppose that the numbers for ANUBIS should be of a similar order.
However, without more concrete input from the experimental side, we are unable to make more definite predictions in this paper.

\begin{figure}\
	\includegraphics[scale=0.65]{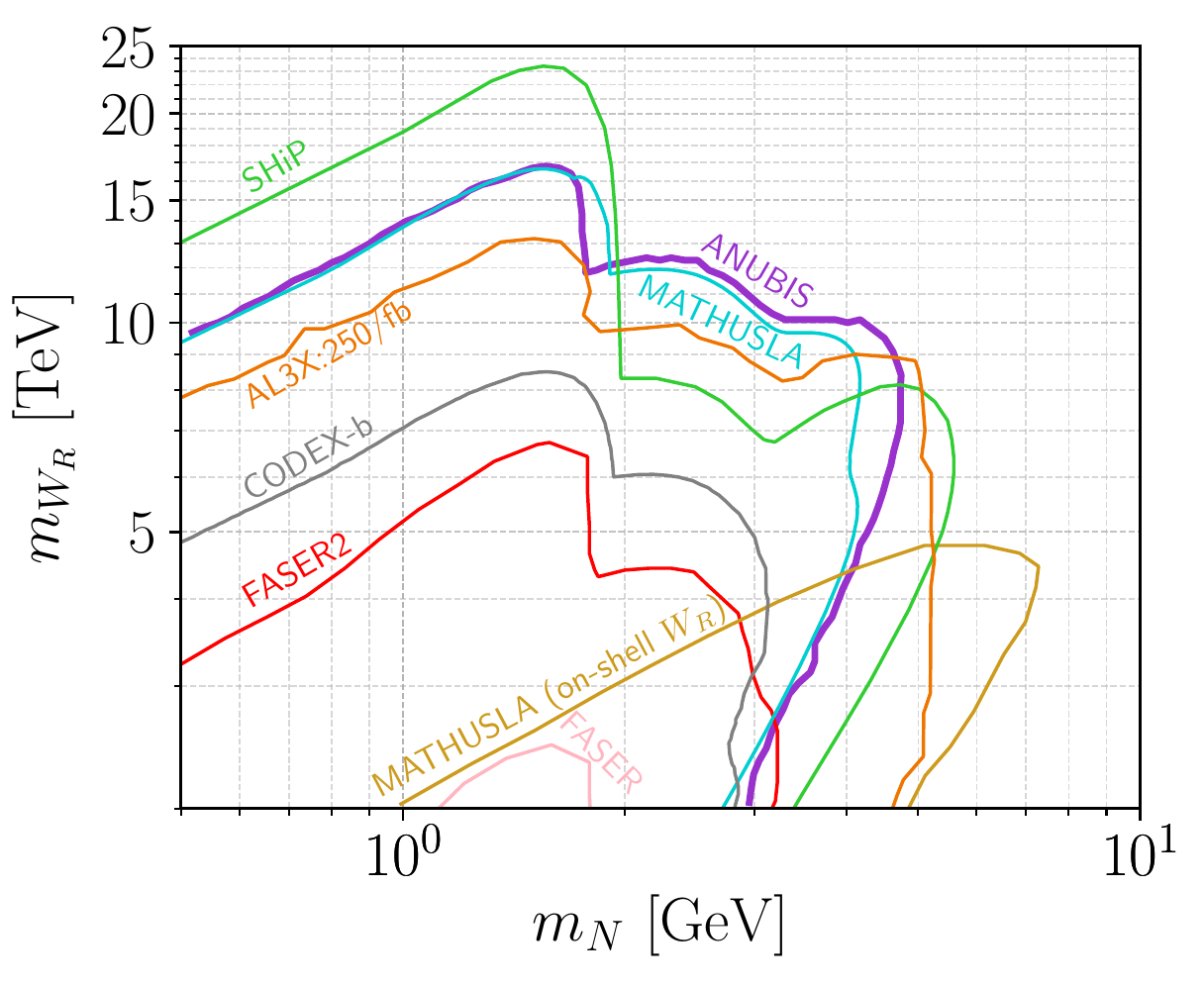}
	\caption{The sensitivity reach of ANUBIS compared with other
		future experiments in the context of the minimal left-right
		symmetric model. All the curves are for HNLs produced from
		bottom and charm mesons decays except one MATHUSLA limit on
		the lower right which is for on-shell $W_R$ decays.}
	\label{fig:HNL_sensitivities_LRSM}
\end{figure}

For the HNLs produced from $B-$ and $D-$mesons decays via an
off-shell $W_R$ at ANUBIS, we may re-interpret the corresponding
results in the context of the minimal left-right symmetric model, by
making the simple substitution $|V_{\alpha N}|^2\rightarrow \Big(
m_{W_L}/m_{W_R} \Big)^4$, since we assume $V^R_{\alpha
	N}\sim\mathcal{O}(1)$ and $g_L/g_R=1$, as discussed in
Sec.~\ref{sec:models}.  We perform the same substitution on the HNL
sensitivities of SHiP, MATHUSLA, AL3X, CODEX-b, FASER, and FASER2 as
well, as their physics reaches are also dominated by the mesons
channels.  We refrain from showing the ATLAS expectations as they are
sensitive to larger values of $m_N$.  We further extract the
MATHUSLA sensitivity for on-shell $W_R$ decays at the HL-LHC from
Ref.~\cite{Curtin:2018mvb} considering the Keung-Sejanovi\'c process
\cite{Keung:1983uu}. The results are presented in
Fig.~\ref{fig:HNL_sensitivities_LRSM} in the plane $m_{W_R}$
vs. $m_N$.  We find that for $m_N$ below the bottom meson threshold
the heavy meson decays to HNLs provide the strongest probe for
$m_{W_R}$, while the limits coming from on-shell $W_R$'s are 
restricted to $m_{W_R} \le 5$ TeV.

\subsection{Results-HNLs from SM-like Higgs decays in $U(1)_{B-L}$ model}

\begin{figure}
	\includegraphics[scale=0.65]{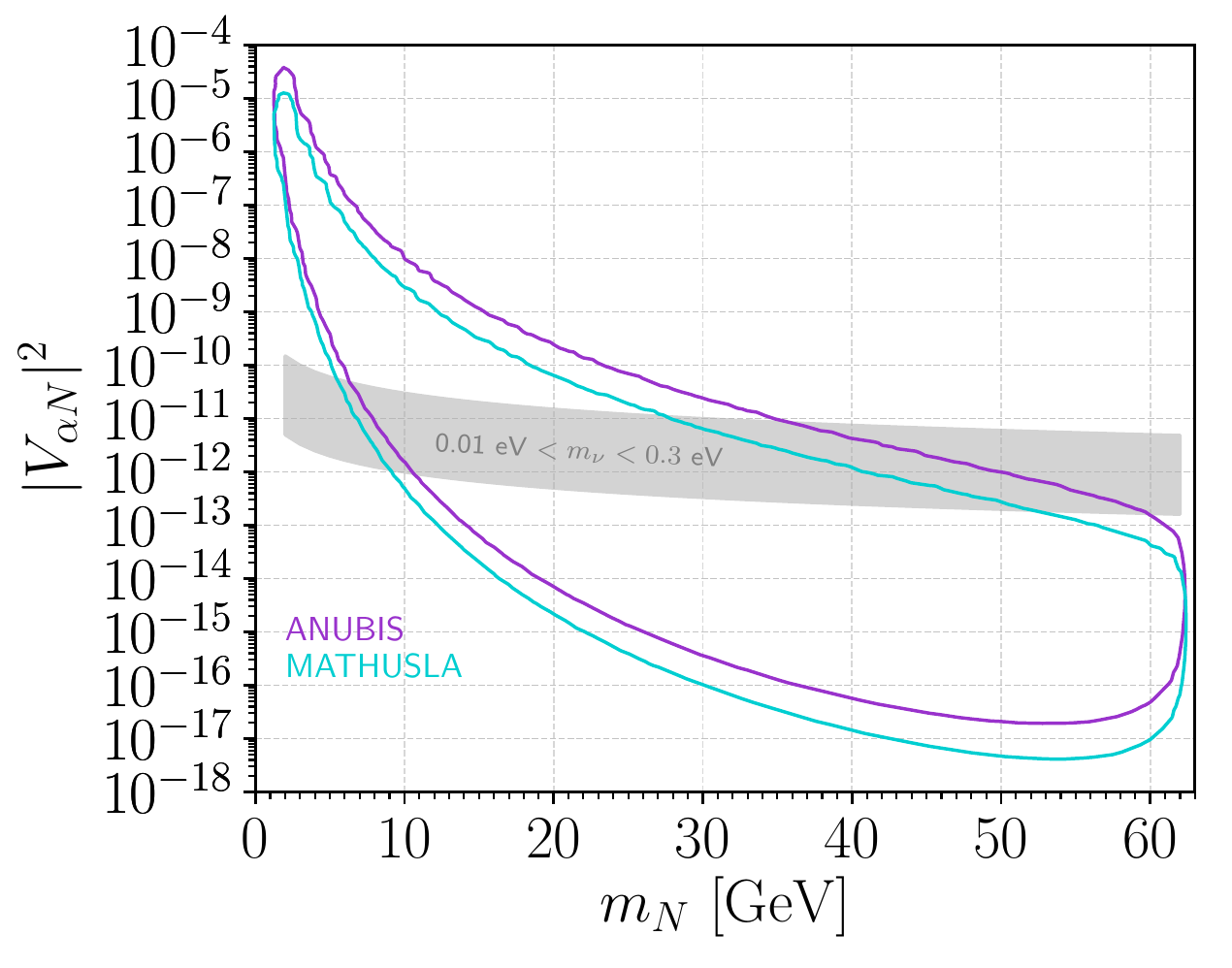}
	\caption{The sensitivity reaches of ANUBIS and MATHUSLA for
		HNLs produced from the SM-like Higgs decays in the context of the
		$U(1)_{B-L}$ model.}
	\label{fig:HNL_sensitivities_U1BmL_Higgs}
\end{figure}

We present the results for another non-minimal HNL scenario with a new
$U(1)_{B-L}$ gauge group and focus on the SM-like Higgs decay into a
pair of HNLs which mix with $\nu_{e/\mu}$.  With the selected
benchmark parameters given in Eq.~\eqref{eqn:U1BmLHiggsParameters}, we
obtain the sensitivities in the plane $|V_{\alpha N}|^2$ vs. $m_N$,
shown in Fig.~\ref{fig:HNL_sensitivities_U1BmL_Higgs}.  The gray band
indicates the interesting area of the parameter space corresponding to the
type-I seesaw limits for light neutrino masses between 0.01 eV and 0.3
eV.  The calculation of the decay probability of HNLs inside MATHUSLA
follows from applying the formulas given in Ref.~\cite{Dercks:2018eua}
for the $200$ m $\times 200$ m $\times 20$ m setup.  We find that for
$m_N$ between 5 (6) GeV and 52 (60) GeV, MATHUSLA (ANUBIS) may probe
the type-I seesaw limits on $|V_{\alpha N}|^2$ for the considered
range of $m_{\nu}$. Though the two detectors show similar sensitivity
reaches, compared to MATHUSLA ANUBIS is sensitive to slightly larger
values of $|V_{\alpha N}|^2$ mainly because of its shorter distance
from the IP.  The sensitive area in the parameter space presented here
may shrink if in the future the experimental upper (lower) bound on
the scalar mixing angle $\beta$ (scalar VEV $\tilde{x}$) gets more
constrained.

\section{Conclusions}\label{sec:conclusions}

Searches for new physics in the form of long-lived particles have
become one of the most studied objectives in the high-energy physics
community in the past few years.  While many BSM models predict new
particles with long lifetimes, heavy neutral leptons, with the 
possibility to explain the non-vanishing active neutrino masses in a
natural way, are perhaps among the most frequently investigated
scenarios for LLPs.

In this work we have studied the sensitivity reach for GeV-scale HNLs
for a recently proposed external detector at ATLAS or CMS, named
`ANUBIS' \cite{Bauer:2019vqk} by performing a Monte-Carlo simulation.
We considered not only the minimal HNL scenario where both production
and decay of the HNLs are controlled by the active-sterile neutrino
mixing, but also extended theoretical scenarios: (i) a left-right
symmetric model and (b) a $U(1)_{B-L}$ model.  In the minimal
scenario, ANUBIS shows similar sensitivities in $|V_{\alpha N}|^2$
($\alpha =e/\mu$) as MATHUSLA for $m_N\lsim 4$ GeV, but has slightly
larger HNL mass reach, thanks to a larger contribution from $W-$boson
decays.  We have recast the results of the minimal scenario for HNLs
produced from heavy mesons decays into the parameter space of the LRSM
($W_R$ vs. $m_N$) where both the production and decay of the HNLs
proceed via an off-shell $W_R$, and have compared ANUBIS with other
experiments.  Fig.~\ref{fig:HNL_sensitivities_LRSM} shows that for
$m_N$ below the $B-$meson threshold, the future experiments SHiP,
ANUBIS, and MATHUSLA would have the largest reach in $m_{W_R}$.

Finally, we estimate the exclusion limits of ANUBIS and MATHUSLA for
the $U(1)_{B-L}$ extension of the Standard Model. For the HNL
production, we focus on the SM-like Higgs boson decaying into a pair
of HNLs.  In Fig.~\ref{fig:HNL_sensitivities_U1BmL_Higgs} we show that
ANUBIS and MATHUSLA have similar sensitivity reaches in the plane
$|V_{\alpha N}|^2$ vs. $m_N$, and both may cover a large part of the
parameter space corresponding to the type-I seesaw predictions based
on the active neutrino masses between 0.01 eV and 0.3 eV.

We conclude that in general ANUBIS is comparable to MATHUSLA in
sensitivity for probing long-lived HNLs in different models, despite
its decay volume being much smaller.  MATHUSLA is more sensitive to
exotic physics with larger values of $c\tau$, as a result of its
larger distance from the IP, which, however, leads only to little
advantage in concrete HNL models.

\medskip

\bigskip
\centerline{\bf Acknowledgements}

\bigskip
M.~H. is supported by the Spanish grant FPA2017-85216-P (AEI/FEDER,
UE), PROMETEO/2018/165 (Generalitat Valenciana) and the Spanish Red
Consolider MultiDark FPA2017-90566-REDC.  Z.~S.~W. is supported by the
Ministry of Science, ICT \& Future Planning of Korea, the Pohang City
Government, and the Gyeongsangbuk-do Provincial Government through the
Young Scientist Training Asia-Pacific Economic Cooperation program of
APCTP.

\bigskip

\bibliographystyle{h-physrev5}

\end{document}